\documentclass[preprint, preprintnumbers, pra,amsmath,amssymb]{revtex4}

\usepackage{graphicx}
\usepackage{dcolumn}
\usepackage{bm}

\begin{document}

\title{Complexity analysis of quantum walk based search algorithms}

\author{B L Douglas}
\author{J B Wang}

\affiliation{School of Physics, The University of Western Australia, 6009, Perth, Australia.}

\begin{abstract}

We present several families of graphs that allow both efficient quantum walk implementations and efficient quantum walk based search algorithms. For these graphs, we construct quantum circuits that explicitly implement the full quantum walk search algorithm, without reference to a `black box' oracle. These circuits provide a practically implementable method to explore quantum walk based search algorithms with the aim of eventual real-world applications.  We also provide a numerical analysis of a quantum walk based search along a twisted toroid family of graphs, which requires O($\sqrt{n}$ log($n$)) elementary 2-qubit quantum gate operations to find a marked node.

\end{abstract}

\maketitle

\section{Introduction}

Search algorithms employing quantum walks have been the focus of several recent studies (see, for example, \cite{Childs2003, Shenvi2003, Childs2004a, Childs2004b, Magniez2007, Santha2008, Reitzner2009}). Along certain types of graphs, 
quantum search algorithms have been shown to yield a quadratic speedup over their classical equivalents, relative to a fixed oracle \cite{Moore2002,Ambainis2005,Potocek2009}. These types of graphs include sparse graphs with efficiently computable neighbours \cite{Aharonov2003}, and highly symmetric graphs such as complete graphs, hypercycles and hypercubes \cite{Reitzner2009,Douglas2009}.

Previous studies of quantum walk based search algorithms have focussed on a computational complexity comparison between the quantum search and the best possible classical search. These search algorithms are usually formulated relative to an oracle, a black box that accepts some input, and returns some corresponding output. A comparison is made as to the relative number of queries to the fixed oracle needed to complete the search. As such, there has been no need to consider the resource requirements of the oracle itself.

However, an efficient practical implementation of such a search algorithm would require an efficiently implementable oracle.  Specifically, it requires the ability to efficiently perform steps of the quantum walk along the graph. So far efficient implementation of quantum walks has been shown to be possible only along some graphs from a certain type, whose global structure is completely characterised by a small number of parameters. Several graphs belonging to this category were considered in \cite{Douglas2009, Loke2011}, and shown to be amenable to exact, efficient quantum circuits implementing quantum walks along them.

Note that there are two differing notions of efficiency considered here. We consider an efficient quantum search algorithm to be one which exhibits at least a quadratic speedup over the best possible classical search, in some oracular setting.  In the case of implementing a quantum walk along a graph on $n$ vertices, we use the term efficient to mean that a step of the walk can be implemented using O(log($n$)) elementary 2-qubit gates. Which of these two situations we are referring to when we mention efficiency should be clear from the context.

In this paper we present examples of families of graphs for which both notions of efficiency hold, that is they yield both efficient quantum walk implementations and efficient quantum walk based search algorithms, without reference to an oracle. For these graphs, we construct quantum circuits that implement a step of the quantum walk (corresponding to an efficient search algorithm) using O(log($n$)) 2-qubit gates, for graphs on $n$ nodes. By specifying explicitly the method by which the walk is implemented, without reference to an oracle, we ensure there are no hidden resource requirements, and as such provide a practically implementable method to experimentally test these search algorithms.

\section{Quantum Walk Based Search}

Quantum walks can be thought of as the quantum analogues of classical random walks. They are a unitary process, and can be naturally implemented by quantum systems (see, for example, references in \cite{Kia2008, Kia2009}. Both discrete time \cite{Tregenna2003,Aharonov2001} and continuous time \cite{Farhi1998,Gerhardt2003} quantum walks can be considered, both of which have yielded several quantum algorithms \cite{Kempe2003, Childs2003, Shenvi2003, Childs2004a, Childs2004b, Ambainis2004, Douglas2008, Berry2010, Berry2011}. Here we consider only discrete time quantum walks. This walk consists of a unitary operator $ U = S C $, acting on the state space, where $S$ and $C$ are termed the shifting and coin operators respectively.

Consider a discrete-time quantum walk along a general undirected graph $G(V,E)$, with vertex set $V=\left\{v_1,v_2,v_3,\ldots\right\}$, and edge set $E=\left\{(v_i,v_j),(v_k,v_l),\ldots\right\}$, being unordered pairs connecting the vertices. The quantum walk acts on an extended position space, in which each node $v_i$, having $d_i$ edges, is split into $d_i$ states, alternatively termed sub-nodes or coin positions. This space then consists of states $\vert v_i , a_i \rangle$, where $1 \le a_i \le d_i$. The shifting operator acts on this extended position space, with its action defined by:
$$S \vert v_i , a_i \rangle = \vert v_j , a_j \rangle \textrm{ , and } S^{2} = \mathbf{I}$$
for some $v_j \in V$ with valency $d_j$, such that $(v_i, v_j) \in E$, and $1 \le a_j \le d_j$. The coin operator comprises a group of unitary operators, or a set of coins, which independently mix the probability amplitudes associated with the group of sub-nodes of a given node. For example, given a vertex $v_i$ with valency $d_i$, the coin can be represented by a unitary $(d_i \times d_i)$ matrix. In the cases we consider here, the coin matrix is always constrained to be symmetric. Berry \textit{et al.} \cite{Berry2011a} has recently developed a software package \textit{qwViz}, which provides an interactive visualisation  of the time-evolution of quantum walks on arbitrarily complex graphs.

One particular algorithmic application of quantum walks is that of searching for a marked node in a graph, and has been the focus of several recent studies (see, for example, \cite{Childs2003, Shenvi2003, Childs2004a, Childs2004b, Magniez2007,Santha2008, Reitzner2009}). These search algorithms generally involve starting in an equal superposition of all possible states, and marking one or more nodes of the graph. The aim is to evolve the state by successive steps of a quantum walk, until the probability to measure the walk at a marked node is sufficiently high, at which point the state of the walk is measured. 

In the following section, we present quantum circuits that implement the full quantum walk based search algorithm.  In this situation the efficiency is not just measured relative to the best possible classical search in an oracular setting. Instead, since in previous quantum search studies \cite{Childs2003,Shenvi2003, Childs2004a, Childs2004b, Magniez2007,Santha2008, Reitzner2009} the oracle is queried once per step of the walk, generally taking the place of the coin or shifting operators of the walk, we will analyse the complexity of these examples based on the number of steps of the quantum walk required, together with the total number of elementary 2-qubit quantum gates required to implement the walk. Explicit circuits implementing the quantum walk based search will be provided for each family of graphs.

\section{Examples}


\subsection{Hypercube}

\begin{figure}[htb]
\includegraphics[width=4.5cm]{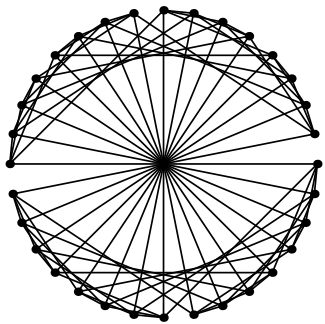}
\caption{\label{hypercube} The 5 dimensional hypercube.}
\end{figure}

Firstly we consider the $n$-dimensional hypercube, with $N=2^n$ nodes, each of degree $n$. The graph for $n=5$ is shown in Fig. \ref{hypercube}. Shenvi \textit{et al.} \cite{Shenvi2003} demonstrated an efficient quantum walk based search algorithm for finding marked nodes on the hypercube. Their algorithm operates quadratically faster than classical methods, requiring O($\sqrt{N}$) calls to an oracle that effectively acts as a coin operator which is biased only relative to the marked node during each step of the quantum walk.

Here we show that this quantum walk oracle can be implemented efficiently, due to the symmetric nature of the graph, requiring O(log($N$)) elementary 2-qubit gates per call. The coin operator $C$ is defined as $C = G \otimes ( \mathbf{I} - \vert x \rangle \langle x \vert) + C^{'} \vert x \rangle \langle x \vert $, where $C'$ is the perturbed coin acting on the marked node $x$ (for the examples presented here, $C' = - \mathbf{I}$). $G$ is the Grover operator, defined on $d$ dimensions by $G_{i,j} = \frac{2}{d} - \delta_{i,j}$. The shifting operator has its standard definition, provided above, such that for $v_i, v_j \in V$,  $S \vert v_i , a_i \rangle = \vert v_j , a_j \rangle$, where $S^{2} = \mathbf{I}$ and $1 \le a_i, a_j \le n$.

\begin{figure}[htb]
\includegraphics[width=7.5cm]{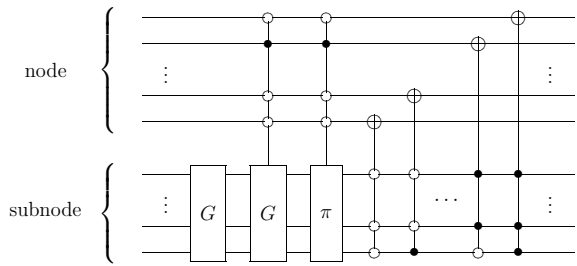}
\caption{\label{circuit_hypercube} Quantum circuit implementing a step of the quantum search along the $n$-dimensional hypercube, on $2^n$ nodes. The circuit contains $n$ `node' qubits and log($n$) `subnode' qubits, and the marked node is represented by the state $\vert 01 \cdots 00 \rangle$.}
\end{figure}

An example quantum circuit implementing a step ($U = S C$) of the walk along the $n$-dimensional hypercube and biased towards the marked node is shown in Fig. \ref{circuit_hypercube}. Here the $G$ operator represents the Grover operator defined above, and the $\pi$ operator (also known as the negative identity operator) represents a $\pi$ phase change applied to each qubit it acts upon. This circuit requires $n$ `node' qubits, with each bit string representing a node of the hypercube, with the natural ordering on the hypercube, such that two nodes are connected if and only if their bit strings have a Hamming distance of 1. Log($n$) `subnode' qubits are required (rounded up to the nearest integer), with the first $n$ bit strings representing the $n$ edges of each node. 

This circuit provides an efficient implementation, in that O(log($N$)) elementary 2-qubit gates are required per step. The Grover operator $G$ acting on the log($n$) subnode qubits requires O(log($n$)) 2-qubit gates \cite{Diao2002}. The generalized conditional NOT operations are described in \cite{Tucci2004}, in which it is shown that a generalized C$^n$NOT gate, performing a NOT operation conditionally based on the state of $n$ other nodes, can be implemented using O($n$) 2-qubit gates together with O($n$) auxilliary qubits. Performing both these and the conditional $G$ and $\pi$ operations requires O($n$) 2-qubit gates and an additional O($n$) auxilliary qubits (see, for example, \cite{Tucci2004,Diao2002}).

\begin{figure}[htb]
\includegraphics[width=7.5cm]{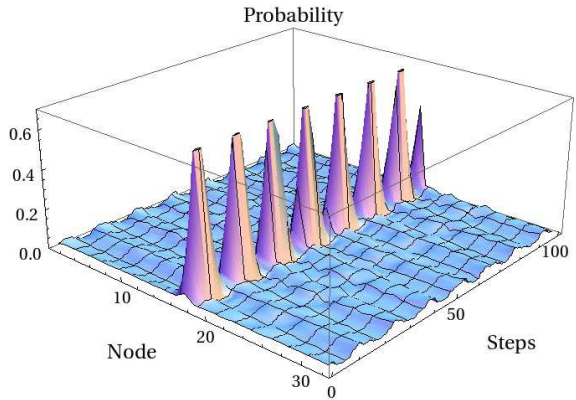}
\caption{\label{hyp_prob} Probability distribution along the hypercube against the number of walking steps.}
\end{figure}

Hence the circuit of Fig. \ref{circuit_hypercube}, performing a step of the walk implementing a quantum search along the hypercube on $n$-dimensions (with $N=2^n$ nodes), can be implemented using O(log($N$)) qubits and O(log($N$)) 2-qubit gates. This circuit is a representation of a single step of the search algorithm of \cite{Shenvi2003}, with the explicit action of the oracle included in the circuit. For the 5-dimensional hypercube, given an initial equal superposition across all states, the resulting probability distribution against number of steps is given in Fig. \ref{hyp_prob}. Hence as in \cite{Shenvi2003}, after O($\sqrt{N}$) steps of the walk (or equivalently O($\sqrt{N}$) repetitions of the circuit of Fig. \ref{circuit_hypercube}, each requiring O(log($N$)) elementary quantum gates), the walk will be found at the marked node with sufficiently high probability.

\subsection{Complete graph}

\begin{figure}[htb]
\includegraphics[width=7.5cm]{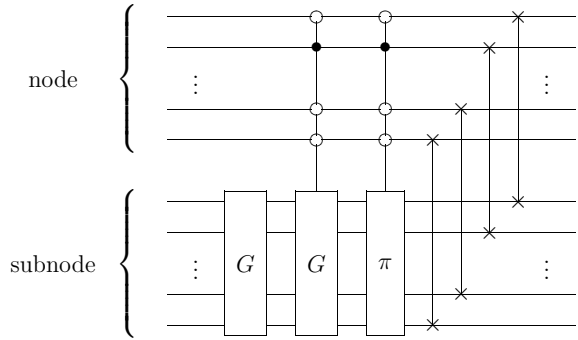}
\caption{\label{circuit_complete} Quantum circuit implementing a step of the quantum search along the complete graph on $2^n$ nodes. The circuit contains $n$ `node' qubits and $n$ `subnode' qubits, and the marked node is represented by the state $\vert 01 \cdots 00 \rangle$.}
\end{figure}

Similar results pertain to the complete graph on $N=2^n$ nodes. Here we consider each node to possess a self loop, so that each node has degree $N$. Combining the circuits of \cite{Douglas2009} for complete graphs with $2^n$ nodes (which efficiently implements a step of the walk along a complete graph) with the search algorithm of \cite{Reitzner2009} yields a complete implementation of the search for the marked node in such graphs. Fig. \ref{circuit_complete} presents this implementation, for a single step of the walk. The shifting and coin operations are applied to the topology of the complete graph exactly as with the hypercube. As in the hypercube example above, for a complete graph on $N=2^n$ nodes, O($n$) node qubits are required. In this case no particular mapping between the bit strings and the nodes is necessary. Since each node has degree $N$, O($n$) subnode qubits are also required, together with O($n$) auxilliary qubits to implement the C$^n G$ and C$^n \pi$ gates. Hence O($n$) = O(log($N$)) elementary 2-qubit gates are again sufficient to implement a step of the walk along this family of graphs. The marked node is found with sufficiently high probability after O($\sqrt{N}$) steps (see \cite{Reitzner2009}), hence in total it requires O($\sqrt{N}$ log($N$)) 2-qubit quantum gates to locate the marked node.

\subsection{Twisted toroid}

\begin{figure}[htb]
\includegraphics[width=7.5cm]{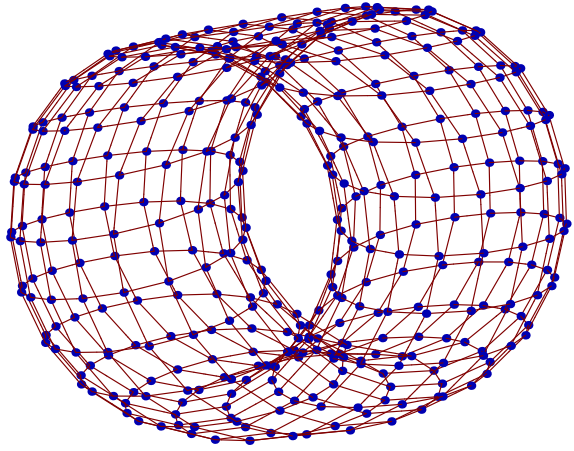}
\caption{\label{toroid_graph} A twisted toroid of dimension $20 \times 20$.}
\end{figure}

Following on from the discussion of the twisted toroid in \cite{Douglas2009}, we considered the possibility of marked node searching along this family of graphs. Formally, a graph in this family of dimension $n \times m$ is constructed by taking an $n \times m$ grid, and associating the endpoints with each other as in the construction of a simple toroid, with the modification : $\{1,j\} \leftrightarrow \{n,(j+1) \textrm{ mod } m\}$ and $\{i,1\} \leftrightarrow \{(i+1) \textrm{ mod } n,m\}$, where `$\leftrightarrow$' denotes an edge. Here $\{i,j\}$ refers to the corresponding point on the grid, where $1 \leq i \leq n, 1 \leq j \leq m$. Fig. \ref{toroid_graph} shows a member of this family of graphs, of dimension $10 \times 10$.

\begin{figure}[htb]
\includegraphics[width=14cm]{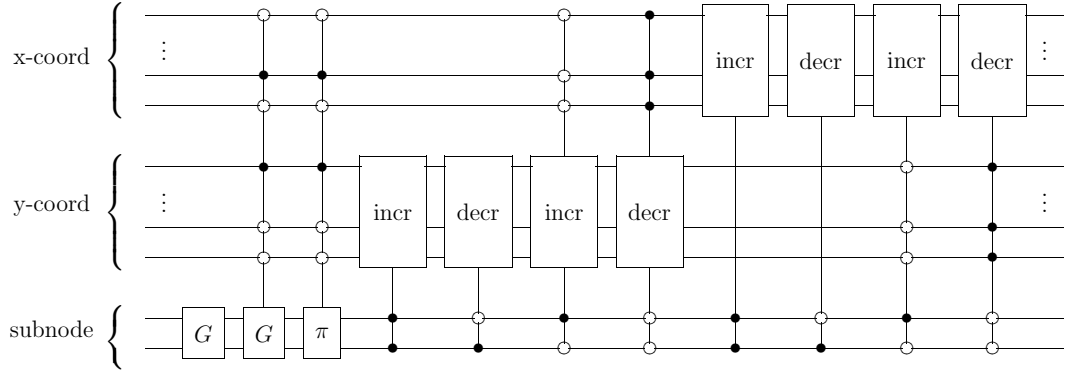}
\caption{\label{circuit_toroid} Quantum circuit implementing a step of the quantum search along the twisted toroid of dimension $2^n \times 2^m$. The circuit contains $n$ `x-coord' qubits and $n$ `y-coord' qubits, and the marked node is represented by the state $\vert 0 \cdots 10,1 \cdots 00 \rangle$.}
\end{figure}

This family of graphs is highly symmetric, and can be completely characterized by a small number of parameters (growing logarithmically with the size of the graph). Given this symmetry, it is a good candidate for a graph on which both quantum search and a direct implementation of quantum walks are efficient. We previously showed in \cite{Douglas2009} that the latter criteria for efficiency holds, in that for a graph of size $n$ a step of an unbiased walk requires O(log($n$)) elementary quantum gates. Fig. \ref{circuit_toroid} extends this result, providing a circuit explicitly performing a step of a search for a marked node, given a twisted toroid of dimension $2^n \times 2^m$.

For this quantum search, the coin and shifting operator are again defined as in the hypercube example above, with the coin operator applying the Grover coin to every node except the marked node, on which the negative identity operator is applied. Each node has degree 4, and the `node' states are represented by bit strings comprising the x- and y-coordinates of the grid from which the twisted toroid was constructed, such that the node $(i,j)$ is represented by the bit string $(i-1)$ concatenated with $(j-1)$. Hence the shifting operator applied to node $(i,j)$ can be implemented by either incrementing or decrementing the bit string associated with each of $i$ and $j$ (modulo $2^n$ and $2^m$ respectively), depending on the coin state of this node.

\begin{figure}[htb]
\includegraphics[width=7cm]{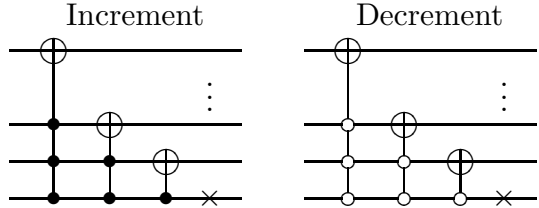}
\caption{\label{incr_decr} Increment and decrement gates on $n$ qubits, producing cyclic permutations of the $2^n$ bit string states. Each C$^i$NOT gate can be implemented using O($i$) 2-qubit gates, hence O($n^2$) 2-qubit gates are sufficient to implement the increment/decrement gates on $n$ qubits.}
\end{figure}

Incrementing and decrementing these bit strings is performed by the `incr' and `decr' operators in Fig. \ref{circuit_toroid}, each of which can be implemented using O($n^2$) 2-qubit gates via the circuits of Fig. \ref{incr_decr}. We are now in a position to calculate an upper bound on the number of 2-qubit gates required to implement Fig. \ref{circuit_toroid}. For a twisted toroid of dimension $2^n \times 2^m$, with $N=2^n \times 2^m$ nodes, $n$ `x-coord' and $m$ `y-coord' qubits are required, together with 2 `subnode' qubits. The controlled G and $\pi$ gates are implementable using O($n+m$) 2-qubit gates, and since the increment and decrement gates on $i$ qubits require O($i^{2}$) 2-qubit gates, the controlled increment and decrement operators are implementable using O($n^2 m + m^2 n$) 2-qubit gates. Note that since the degree of the graph is fixed at 4, only 2 subnode qubits are required regardless of the values of $n$ and $m$. Hence a step of the quantum walk along a twisted toroid on O($N$) nodes can be implemented using O(log($N$)) 2-qubit gates.

\begin{figure}[htb]
\includegraphics[width=7cm]{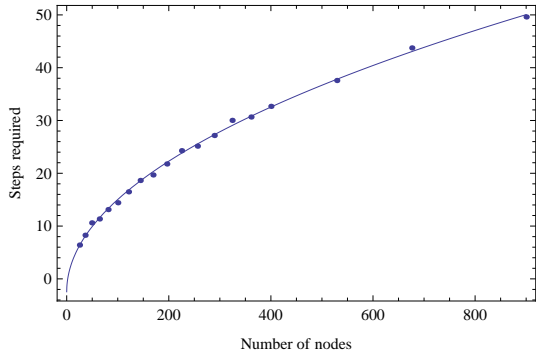}
\caption{\label{toroid_prob} Number of walking steps required to reach maximum success probability against the size of the graph. The solid line represents a $\sqrt{N}$ fit to the numerical data.}
\end{figure}

It remains to show whether or not the circuit of Fig. \ref{circuit_toroid} can perform an efficient search for the marked node - i.e. whether for a twisted toroid of dimension $n \times m$, the marked node can be found with O($\sqrt{n} \times \sqrt{m}$) iterations of this circuit. To examine this, the behaviour of the walk was analyzed for a range of different toroid sizes. Regardless of the size, the search behaviour was in one respect successful, in that the probability to find the walk at the marked node reached a sufficiently high level after a certain number of steps, with fixed periodic behaviour. The approximate number of steps required was found for a range of twisted toroid sizes by calculating the period of the success probability (to find the walk at the marked node) with number of steps. The resulting correlation between toroid size (in number of vertices) and approximate number of steps required to reach maximum success probability is shown in Fig. \ref{toroid_prob}. These numerical results show that for a twisted toroid with $N$ nodes, O($\sqrt{N}$) steps of the walk are required to find the marked node, matching (up to a constant factor) the theoretical lower bound for unstructured search \cite{Bennett1997}.

Putting these results together, we see that this family of twisted toroids satisfy both criteria for efficient search via quantum walks, yielding for a graph on $N$ nodes both an efficient implementation of the walks themselves (using O(log($N$)) 2-qubit gates), and requiring O($\sqrt{N}$) steps of the walk to find a marked node with sufficiently high probability.

\section{Conclusions}

The examples given in this paper are intended to provide a `proof of concept', presenting graphs for which both efficient quantum search and efficient implementation of quantum walks are possible. By explicitly characterising the internal action of the oracle, these circuits provide a practically implementable method to experimentally test these quantum walk based search algorithms. In some previous studies, quantum speedups have been shown relative to oracles that cannot be efficiently implemented. Whilst these studies are very useful from a computational complexity viewpoint, the algorithms involved cannot be practically implemented.

\newpage

\bibliography{quantum_walk_search}

\end{document}